\documentclass[times]{jtitauth}
\usepackage[T1]{fontenc}
\usepackage{authblk}
\usepackage{indentfirst}
\usepackage{amsmath,amssymb,amsfonts}
\usepackage{algorithmic}
\usepackage{multirow}
\usepackage{footnote}
\usepackage[symbol]{footmisc}

\begin{document}
\title{Energy Consumption in Wireless Systems Equipped with RES, UAVs, and IRSs}

\author{Adam Samorzewski*}
\affil[]{*\textit{Institute of Radiocommunications, Poznan University of Technology, Poznan, Poland}}
\markboth{Adam Samorzewski}{Energy consumption in wireless systems equipped with RESs, UAVs, and IRSs}

\maketitle

\begin{abstract}
The paper considers the characteristics of the energy budget for mobile base stations (BSs) in the form of Unmanned Aerial Vehicles (UAVs) equipped with Radio Frequency (RF) transceivers, Intelligent Reconfigurable Surfaces (IRSs), and Renewable Energy Sources (RESs). The obtained results highlight the benefits and challenges related to using the aforementioned mobile base stations from the energy side. The research cases took into account two types of UAV devices -- multirotor and fixed-wing (airplane-like).\footnote[7]{The work has been realized within research project no. 2021/43/B/ST7/01365 funded by the National Science Center in Poland.}\footnote[8]{Copyright \copyright\text{ } 2023 National Institute of Telecommunications in Poland. Personal use is permitted. For any other purposes, permission must be obtained from the National Institute of Telecommunications in Poland by emailing journal@jtit.pl. This is the author's version of an article that has been published in the \textit{Journal of Telecommunications and Information Technology (JTIT)} by the National Institute of Telecommunications in Poland. Changes were made to this version by the publisher before publication, the final version of the record is available at: https://dx.doi.org/10.26636/jtit.2023.170923. To cite the paper use: A.~Samorzewski, "Energy Consumption in Wireless Systems Equipped with RES, UAVs, and IRSs," \textit{Journal of Telecommunications and Information Technology (JTIT)}, 2023, no. 2, pp.~35--40, doi: 10.26636/jtit.2023.170923 or visit https://jtit.pl/jtit/article/view/1175/1161.}
\end {abstract}


\section{Introduction}
\label{secIntroduction}
In recent years, obtaining energy from solar radiation has become a very common solution that finds application not only in industry but also in many different areas of everyday life. The process of obtaining solar energy is carried out using generators in the form of so-called photovoltaic panels (PV -- Photovoltaics). The current market attractiveness of photovoltaics results not only from the ecological approach to obtaining energy resources but also from the prices of solar modules, which have been on a downward trend for the last few years \cite{1}. Therefore, it is not surprising that photovoltaic panels are now also used in telecommunications systems. Particular benefits can be seen from the perspective of heterogeneous wireless systems such as mobile networks.

Wind energy has also gained momentum.~Generating energy from wind power is a process that depends mainly on the location, current weather conditions, and altitude above sea level.~A generator that extracts energy from the wind is called a wind turbine (WT). A wind turbine converts the flowing wind speed into mechanical energy and then into electrical energy. Wind energy can be a good alternative to solar energy, and for some locations or seasons, it may turn out to be an even more reliable solution \cite{2}, \cite{3}.

Current wireless systems are powered mainly by conventional energy sources, the utilization of which is still progressing. Due to the increasing transmission (service) requirements of Information and Communication Technology (ICT) networks, with each subsequent year, they show an increasing energy demand compared to previous years. Equipping wireless network access points with Renewable Energy Sources (RESs) could reduce or completely eliminate the demand for electricity from conventional sources.~This, in turn, would not only reduce carbon dioxide $\left(\text{CO}_{2}\right)$ emissions to the atmosphere but would also slow down the consumption of non-renewable resources.~However, despite the benefits of RESs, they also have one basic disadvantage -- the supply of energy resources is not guaranteed in an even and continuous manner due to changing weather conditions.~Therefore, it may be necessary to engage more than one type of renewable source at the same time or to develop algorithms to optimize the use of available energy \cite{2}--\cite{4}.

In addition, one of the paradigms in present and future telecommunications systems is the provision of services in hard-to-reach areas.~Such areas are characterized by unfavorable conditions for the development of telecommunications infrastructure, which result, for example, from the terrain or the inability to connect to the existing telecommunications and energy networks. In a situation where the first problem could be solved by using mobile access nodes (e.g., drones) and involving wireless connectivity for backhaul communication, the second case is extremely difficult to cope with \cite{5}.~However, considering the off-grid approach in the context of powering individual system components, RES generators (e.g., photovoltaic panels, wind turbines) in combination with an autonomous battery system could be able to meet the energy needs of the Radio Access Network (RAN) \cite{6}. However, this requires appropriate modeling of the energy cycle of the system, i.e., predicting the results of production and the use of energy resources in time in such a way as to ensure their adequate amount for each type of load on network links \cite{7}.

In the context of increasing the efficiency of wireless systems, the Intelligent Reflective Surface (IRS) is proving to be a very promising concept.~The IRS is a device in the form of a flat surface containing a large number of passive reflecting elements.~Each of these elements is able to independently cause a controlled change in the amplitude and/or phase of the incident radio signal. The dense distribution of IRS surfaces in the wireless network and intelligent coordination of their reflections can make the propagation of the radio signal (wireless channels) between transmitters and receivers flexibly reconfigurable.~This, in turn, can solve the problems of wireless channel dropout and interference. In addition, thanks to the use of IRS, improvements in the throughput and reliability of wireless communication can be achieved \cite{8}.

This paper proposes models of energy consumption for wireless access networks whose nodes are Unmanned Aerial Vehicles (UAVs) equipped with Intelligent Reconfigurable Surfaces (IRSs) and powered by Renewable Energy Sources (RESs).~In the contribution, the characteristics of the energy balance for the aforementioned system depending on the type of equipment and weather conditions have been also investigated.

The paper has been organized as follows: first, the mathematical models for energy consumption and harvesting have been described; next, the simulation setup configuration has been presented.~After the third section, the results of performed simulation runs have been overviewed.~Finally, the last part is focused on conclusions summarizing the contribution of the paper.

\section{Energy Models}
\label{secEnergyModels}
In order to estimate the energy efficiency of the considered wireless system, an extremely important step is to precisely configure the study scenario and to model the energy cycle for the components of the network.~The simulation scenario considered in the paper assumes the use of a~UAV device as a base station (BS), which is equipped with photovoltaic panels and wind turbines for energy generation. In addition, the mobile access node has one IRS device to extend the range of the emitted radio signal and improve the performance of the wireless system.~The used UAV base station may be of the multirotor or fixed-wing type, which will allow comparing the energy balance characteristics for both kinds. The UAV node is able to maintain a connection with the major station on the backhaul link implemented in accordance with microwave (MW) technology and on the access link realized in the Radio Frequency (RF) band.~The UAV base station hovers/flies above the ground in free space in the city of Poznan.~Where a multirotor mobile station can maintain a~fixed position in the sky, a fixed-wing type station is forced to follow some flight trajectory. The work considers a static case in which drones do not change their position (multirotor) or trajectory (fixed-wing) -- optimization of the location of the UAV station and its impact on the functioning of the network is the subject of parallel activities.~However, the paper takes into account the impact of atmospheric factors such as wind or cloud cover on both the consumption of energy resources and their harvesting from RESs.~The power consumption (PC) characteristics take into account the utilization related to the hovering/flying of the UAV station and the provision of services by it in the considered area using the MIMO technique and IRS device. However, in the context of obtaining energy resources from RESs, the generation processes were analyzed both for different times of the day and year, as well as for different wind speeds, solar radiation densities, and types of cloud cover, i.e., different cloud thicknesses.~Telecommunications traffic in the network is assumed to be fixed.

\subsection{Energy Consumption}
\label{subSecConsumption}
The power consumption for a fixed-wing UAV mobile base station $\left(P_\text{UAV,FW}\right)$ in time step $t$ can be described by the following formula \cite{9}:
\begin{align}
    \label{eqFixedWing}
    &P_\text{UAV,FW}\left(t\right)=\\ \nonumber &\left|c_{1} v_\text{UAV}^{3}\left(t\right)+\frac{c_{2}}{v_\text{UAV}\left(t\right)}\left(1+\frac{a_{\perp}^{2}\left(t\right)}{g^{2}}\right)+m_\text{ALL}a_{\parallel}\left(t\right)v_\text{UAV}\left(t\right)\right|,
\end{align}
where $v_\text{UAV}$ is the forward flight velocity of the UAV, $a_{\perp}$~and $a_{\parallel}$ are the centripetal acceleration when the base station moves along the circular trajectory, and the forward acceleration, respectively.~The $m_\text{ALL}$ is the sum of the masses of all the components of a particular UAV, i.e., the weights of the UAV itself $\left(m_\text{UAV}\right)$, its battery system $\left(m_\text{BATT}\right)$, RES generators $\left(m_\text{PV},\text{ } m_\text{WT}\right)$, and additional equipment $\left(m_\text{RF},\text{ } m_\text{IRS},\text{ } m_\text{PKG}\right)$. Next, the $g$ parameter is the gravitational acceleration, and $c_{1}$ and $c_{2}$ are the parameters described by the following formulas \cite{9}:
\begin{align}
    \label{eqC1C2}
    c_{1}\overset{\Delta}{=}\frac{1}{2}\rho C_{\text{D}_0}S, & \qquad c_{2}\overset{\Delta}{=}\frac{2W^{2}}{\pi e_{0} A_\text{R}\rho S},
\end{align}
where $\rho$ is the air density (calculated according to the formula from \cite{10}), $C_{\text{D}_0}$ is the zero-lift drag coefficient and $S$~is the reference area. The parameter of $W=m_\text{ALL}\cdot g$ is the weight force of the UAV device.~Whereas $e_{0}$ and $A_\text{R}$ are wingspan efficiency (Oswald efficiency) and UAV wing aspect ratio, respectively.

As for the power consumption of multirotor base stations $\left(P_\text{UAV,MR}\right)$ in the current time step $t$, its mathematical model is as follows \cite{11}:
\begin{align}
    \label{eqMultirotor}
    \nonumber
    &P_\text{UAV,MR}\left(t\right)=\frac{d_{0}}{2}\rho n s A_\text{UAV}v_\text{UAV}^{3}\left(t\right)+P_{0}\left(t\right)\left(1+\frac{3v_\text{UAV}^{2}\left(t\right)}{\Omega^{2}\left(t\right)R^{2}}\right)\\ &+P_\text{i}\tilde{\kappa}\left(t\right)\left(\sqrt{\tilde{\kappa}\left(t\right)+\frac{v_\text{UAV}^{4}\left(t\right)}{4v_{0}^{4}}}-\frac{v_\text{UAV}^{2}\left(t\right)}{2 v_{0}^{2}}\right)^{\frac{1}{2}},
\end{align}
where $\Omega$, $R$, $v_{0}$ are the angular velocity, radius and mean induced velocity in the hover of a single rotor of a UAV, respectively. The latter can be expressed as $v_{0}=\sqrt{\frac{W}{2\rho A_\text{UAV}}}$, where $A_\text{UAV}$ is the area of a rotor of the UAV. The parameters of $n$ and $s$ are the number of UAV rotors and the solidity of a~single one. In turn, $\tilde{\kappa}$ and $d_{0}$ are the ratio of thrust to weight forces of the UAV and the fuselage drag ratio. The values of the parameters of $P_{0}$ $\left(\text{in time step }t\right)$ and $P_\text{i}$ are described by the following formulas \cite{11}:
\begin{align}
    \label{eqP0Pi}
    P_{0}\left(t\right)=\frac{\delta}{8}\rho n s A_\text{UAV} \Omega^{3}\left(t\right) R^{3},\quad P_{i}=\left(1+k\right)\left(\frac{W^{3}}{2 \rho n A_\text{UAV}}\right),
\end{align}
where $\delta$ is the profile drag coefficient and $k$ is the incremental correction factor to induced power.

The power utilized by the IRS device connected to the UAV $\left(P_\text{IRS}\right)$ and RF transceiver $\left(P_\text{MIMO}\right)$ has been evaluated in accordance with the mathematical models contained in \cite{12} and \cite{13}, respectively. Those models are described by the equations (\ref{eqIRS}) and (\ref{eqMIMO}) attached below:
\begin{align}
    \label{eqIRS}
    P_\text{IRS}\left(t,b\right)=N\left(t\right)\cdot P_\text{n}\left(b\right),
\end{align}
\begin{align}
    \label{eqMIMO}
    &P_\text{MIMO}\left(t\right)=P_\text{FIX}+P_\text{TC}\big(M\left(t\right)\big)+P_\text{CE}\big(M\left(t\right),K\left(t\right)\big)\\ \nonumber &+P_\text{C/D}\big(\text{TR}_\text{UL}\left(t\right),\text{TR}_\text{DL}\left(t\right)\big)
    +P_\text{BH}\big(\text{TR}_\text{UL}\left(t\right),\text{TR}_\text{DL}\left(t\right)\big)\\ \nonumber &+P_\text{SP}\big(M\left(t\right),K\left(t\right)\big)+\frac{P_\text{TX}\left(t\right)}{\mu_\text{PA}},
\end{align}
where $P_\text{FIX}$ is the fixed power demand of a particular mobile base station, $P_\text{TC}$ is the consumption of transceiver chains, and $P_\text{CE}$, $P_\text{C/D}$, and $P_\text{SP}$ are the power components utilized for operations performed within a base station such as channel estimation, data coding/decoding, and signal processing. Next, $P_\text{BH}$ and $P_\text{TX}$ are the powers related to mobile data flow, i.e., load-dependent MW backhaul and RF fronthaul, respectively. The parameters of $M\left(t\right)$ and $K\left(t\right)$ are the numbers of active antenna elements and served users in the current time step $t$. Next, $\mu_\text{PA}$ is the efficiency of the power amplifier. Finally, the $\text{TR}_\text{UL}$ and $\text{TR}_\text{DL}$ components are uplink (UL) and downlink (DL) total data throughputs, which can change over time. In addition, for channel estimation and signal processing the use of the minimum mean-squared error (MMSE) scheme has been assumed.

Eq.~(\ref{eqIRS}) consists of only two parameters -- $N$ and $P_\text{n}$, which are the number of identical reflecting elements that effectively perform phase shifting on the impinging signal, and the power consumption of each phase shifter. The value of the second one is dependent on the bit resolution $b$ of the used phase shifter.

\subsection{Energy Generation}
\label{subSecGeneration}
Models of harvesting energy resources by the applied generators of Renewable Energy Sources are presented below. The power generated by the photovoltaic panel of the UAV base station $\left(P_\text{PV}\right)$ can be described by the following mathematical formula \cite{14}:
\begin{align}
    \label{eqPV}
    P_\text{PV}\left(t\right)=P_\text{r,PV}f_\text{PV}\left(\frac{\overline{G}_\text{T}\left(t\right)}{\overline{G}_\text{T,STC}}\right)\left[1+\alpha_\text{P}\left(T_\text{c}\left(t\right)-T_\text{c,STC}\right)\right],
\end{align}
where $P_\text{r,PV}$ is the rated capacity of the PV array (its output power under Standard Test Conditions -- STC), $f_\text{PV}$ is the PV derating factor, and $\alpha_\text{P}$ is the temperature coefficient of power, which indicates how strongly the PV array power output depends on the cell temperature. In turn, $\overline{G}_\text{T}$ and $\overline{G}_\text{T, STC}$ are the solar incident radiation on the PV array in the current time step $t$ and at STC, respectively.~Whereas $T_\text{c}$ and $T_\text{c,STC}$ are also the PV cell temperature in the current time step $t$ and under STC, respectively.~The PV cell temperature can be calculated with the equation contained in \cite{14}.

\begin{figure}[h]
\label{figPoweCurve}
\centering
\includegraphics[width=0.5\textwidth]{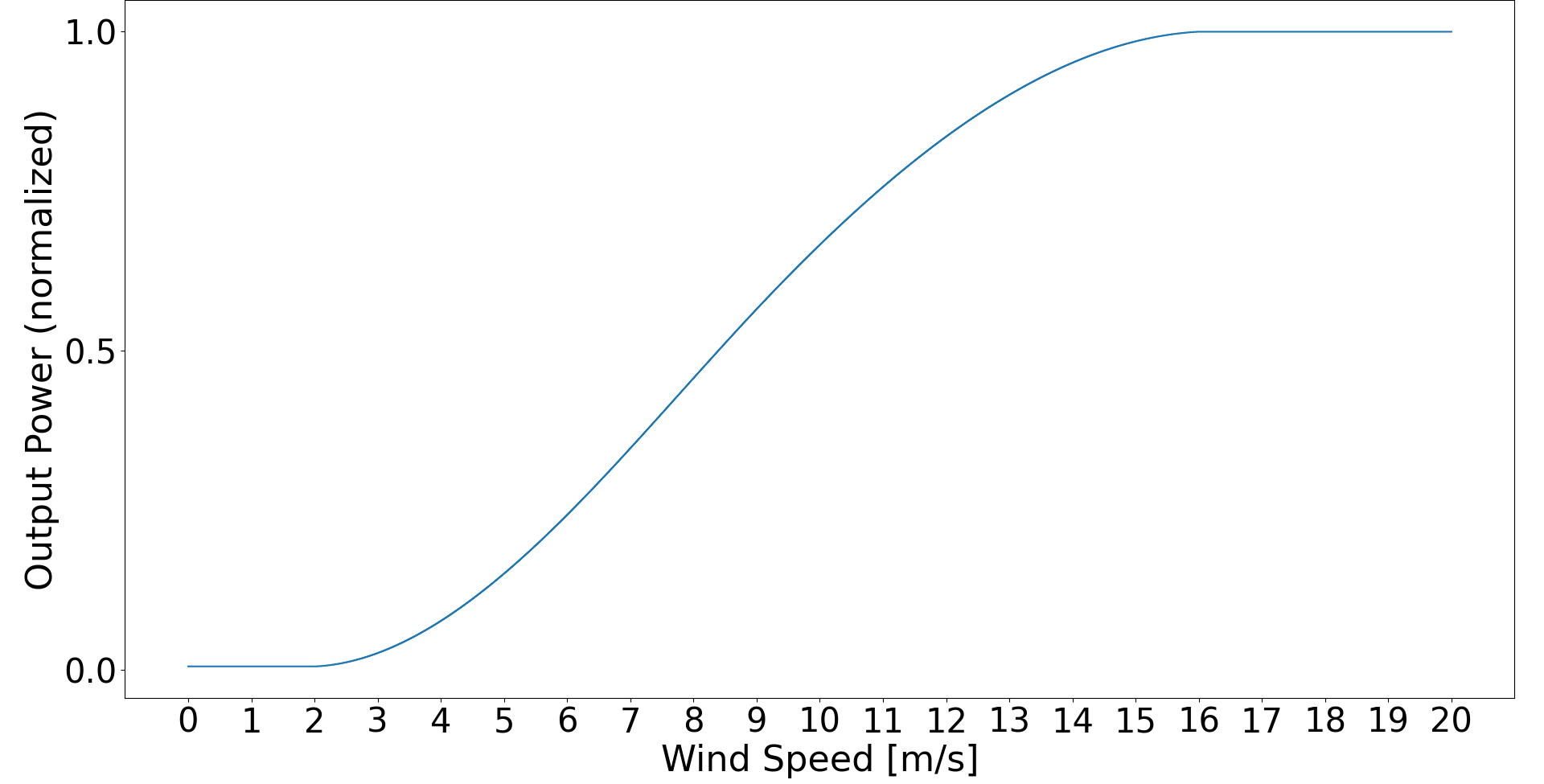}
\caption{Power curve function of wind turbine (based on \cite{16}).}
\end{figure}

The power generated by the wind turbine $\left(P_\text{WT}\right)$ can be described as follows \cite{15}:
\begin{equation}
    \label{eqWT}
    P_\text{WT}\left(t\right)=
    \begin{cases}
        0, & \text{if } v_\text{w}\left(t\right) < v_\text{in} \\
        F\big(v_\text{w}\left(t\right)\big), & \text{if } v_\text{in} \leqslant v_\text{w}\left(t\right) < v_\text{r} \\
        P_\text{r,WT}, & \text{if } v_\text{r} \text{ } \leqslant v_\text{w}\left(t\right) \leqslant v_\text{out},
    \end{cases}
\end{equation}
where $v_\text{in}$, $v_\text{out}$, $v_\text{w}$ are the cut-in and cut-out speed of the wind turbine (defining the range of wind speed values for which the wind turbine is able to generate energy resources), and the instantaneous wind speed, respectively. Next, $F$ is the power curve function of the WT, which determines its energy harvesting characteristic.~Finally, $P_\text{r}$ and $v_\text{r}$ denote the rated output power of the used wind turbine and minimum wind speed $\left(v_\text{in} < v_\text{r} < v_\text{out}\right)$, for which this power can be achieved.~The power curve function normalized in the relation to the maximum value of the WT's output power has been presented in Fig.~\ref{figPoweCurve}.~The corrections of wind speed (to get its correct value at the altitude of the generator) and WT output power (which depends on the current air density) were realized in accordance with \cite{14}.

\section{Simulation Setup}
\label{secSimSetup}
The examined base stations were UAVs of the multirotor and fixed-wing type.~Furthermore, each UAV BS was also equipped with a single RF transceiver and IRS device. The impact of RES generators enabling on the overall power consumption of the mobile base station has been also taken into consideration.~Thus, four study cases have been simulated.~The first investigates PC in different seasons of the year, which is caused by the UAV and its additional equipment (IRS device and RF transceiver) with no RES generators on board. The second, third, and fourth cases take into account as well the power consumption resulting from powering the mobile BS with a PV, WT, and both of them, respectively. The examination of energy generation from RESs has been considered based on the real products, the specifications of which can be found in \cite{16} and \cite{17}.

Simulations have been prepared using the Python programming language.~As the outcomes, the characteristics for energy consumption of the mobile base stations as well as for energy harvesting of the RES generators have been evaluated.~The results have been prepared for four different dates starting various seasons of the last year -- vernal equinox $\left(20^\text{th} \text{ March } 2022\right)$, summer solstice $\left(21^\text{st} \text{ June } 2022\right)$, autumn equinox $\left(23^\text{rd} \text{ September } 2022\right)$, and winter solstice $\left(21^\text{st} \text{ December } 2022\right)$. All the weather data used in the simulations are the real historical data taken from the following source \cite{18} for the city of Poznan (Poland).~The one exception is the global solar radiation density which was estimated by using the MAC model \cite{19}.

\begin{table}[t]
\label{tabSimParams}
\caption{VALUES OF THE SIMULATION PARAMETERS \cite{7},~\cite{9}, \cite{11}, \cite{12}, \cite{16}, \cite{17}, \cite{20}--\cite{23}}
\resizebox{0.485\textwidth}{!}{
\begin{tabular}{llcc}
\hline
\hline
Symbol                           & Quantity                         & Value             & Unit                       \\ \hline
$m_\text{UAV}$                   & mass of a UAV                      & $5.0$               & $\left[\text{kg}\right]$                   \\
$m_\text{BATT}$                  & mass of a battery                  & $0.94$              & $\left[\text{kg}\right]$                   \\
$m_\text{RF}$                    & mass of a RF trans.                & $2.0$               & $\left[\text{kg}\right]$                   \\
$m_\text{IRS}$                   & mass of an IRS dev.                 & $1.0$               & $\left[\text{kg}\right]$                   \\
$m_\text{PV}$                    & mass of a PV panel                 & $2.78$              & $\left[\text{kg}\right]$                   \\
$m_\text{WT}$                    & mass of a WT                       & $6.0$               & $\left[\text{kg}\right]$                   \\
$m_\text{PKG}$                   & mass of an addi. package            & $0.0$               & $\left[\text{kg}\right]$                   \\
$\Omega$                         & rotor angular velocity           & $300.0^{*}$             & $\left[\text{rad}/\text{s}\right]$                \\
\multirow{2}{*}{$v_\text{UAV}$}  & \multirow{2}{*}{velocity of 
a UAV} & $0.0^{*}$  & \multirow{2}{*}{$\left[\text{m}/\text{s}\right]$} \\
                                 &                                  & $10.0^{**}$  &                            \\
$v_\text{in}$                    & cut-in speed of a WT               & $2.0$               & $\left[\text{m}/\text{s}\right]$                  \\
$v_\text{out}$                   & cut-out speed of a WT              & $20.0$              & $\left[\text{m}/\text{s}\right]$                  \\
$v_\text{r}$                     & rated speed of a WT                & $16.0$              & $\left[\text{m}/\text{s}\right]$                  \\
$a_{\parallel}$                  & UAV's forward accel.             & $0.0^{**}$               & $\left[\text{m}/\text{s}^{2}\right]$                 \\
$a_{\perp}$                      & UAV's centripetal accel.         & $0.0^{**}$               & $\left[\text{m}/\text{s}^{2}\right]$                 \\
$g$                              & gravitational acceler.            & $9.81$              & $\left[\text{m}/\text{s}^{2}\right]$                 \\
$\delta$                         & profile drag coefficient         & $0.012^{*}$             &                            \\
$k$                              & incre. correction factor         & $0.1^{*}$               &                            \\
$\tilde{\kappa}$                 & thrust-to-weight ratio           & $1^{*}$                 &                            \\
$d_{0}$                          & fuselage drag ratio              & $14.52^{*}$             &                            \\
$C_{\text{D}_{0}}$                 & zero-lift drag coeffi.           & $0.01^{**}$              &                            \\
$e_{0}$                          & Oswald efficiency                & $0.85^{**}$              &                            \\
$A_\text{R}$                     & UAV's wing aspect ratio                & $118.81^{**}$            &                            \\
$S$                              & UAV's ref. (wing) area            & $1.0^{**}$               & $\left[\text{m}^{2}\right]$                   \\
$A_\text{UAV}$                   & UAV's rotor area                       & $0.071^{*}$             & $\left[\text{m}^{2}\right]$                   \\
$R$                              & UAV's rotor radius                     & $0.15^{*}$              & $\left[\text{m}\right]$                    \\
$s$                              & UAV's rotor solidity                   & $0.067^{*}$             &                            \\
$n$                              & number of rotors                & $8^{*}$                 &                            \\
$M$                              & number of antenna ele.           & $16$                &                            \\
$K$                              & number of users                  & $10$                &                            \\
$N$                              & number of reflect. ele.          & $16$                &                            \\
$b$                              & phase shifter resolution         & $6$                 & $\left[\text{bit}\right]$                  \\
$\text{TR}_\text{UL}$                   & UL data throughput                    & $50.0$              & $\left[\text{Mbps}\right]$                 \\
$\text{TR}_\text{DL}$                   & DL data throughput                    & $100.0$             & $\left[\text{Mbps}\right]$                 \\
$f_\text{PV}$                    & PV's derating factor                  & $0.72$              &                            \\
$\alpha_\text{P}$                & tem. coeffi. of power            & $-0.5$              & $\left[\%/^{\text{o}}\text{C}\right]$                \\
$T_\text{c,STC}$                 & STC solar cell temper.             & $25$                & $\left[^{\text{o}}\text{C}\right]$                   \\
$G_\text{T,STC}$                 & STC solar incident radi.           & $1000$              & $\left[\text{W}\right]$                    \\
$P_\text{r,PV}$                  & PV panel's rated power           & $20.0$              & $\left[\text{W}\right]$                    \\
$P_\text{r,WT}$                  & rated power of a WT                & $30.0$              & $\left[\text{W}\right]$                    \\
$P_\text{n}$                     & power of a phase shifter           & $7.8$               & $\left[\text{W}\right]$                    \\
$P_\text{TX}$                  & transmit power of a UAV                & $15.0$              & $\left[\text{W}\right]$                    \\
$\mu_\text{PA}$                     & power amplifier efficiency           & $0.35$               &                     \\
\hline
\hline
\end{tabular}}
\footnote[1]{*}{\footnotesize \text{ } only for the multirotor type} \qquad \footnote[7]{*}{\footnotesize \text{ } only for the fixed-wing type}
\end{table}

All the prepared energy characteristics charts have been normalized in relation to the maximum values obtained for them. Furthermore, in Tab.~\ref{tabSimParams} the values for the parameters of the mathematical energy models aforementioned in Section \ref{secEnergyModels} have been highlighted. The values of the parameters related to the power consumption model for RF transmission $(\text{not contained in Tab.~\ref{tabSimParams}})$ were taken from Tab.~5.3 in \cite{13}.

\section{Results}
\label{secResults}
Below are the charts with the characteristics of energy consumption by the mobile base stations as well as of energy generation by their RES generators have been shown.~A~particular color of the plots and bars is strictly associated with a specific season of the year: blue -- the vernal equinox, yellow -- the summer solstice, green -- the autumn equinox, and red -- the winter solstice.

\begin{figure}[h]
\label{figConsumption}
\centering
\includegraphics[width=0.49\textwidth]{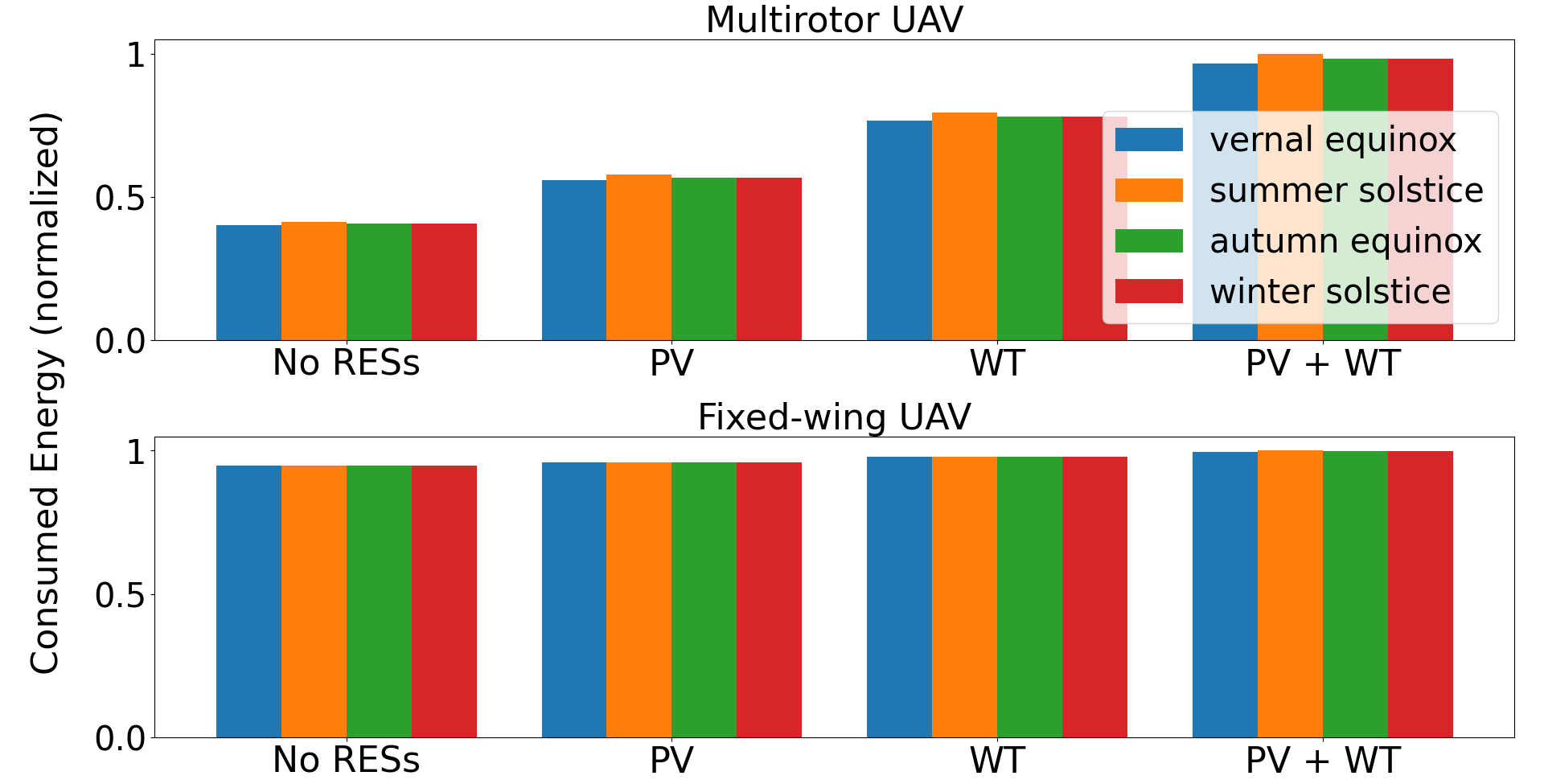}
\caption{Energy consumption character. for multirotor and fixed-wing UAVs.}
\end{figure}

In Fig.~\ref{figConsumption}, the normalized energy utilization process for the different UAV types has been presented.~The chart at the top denotes the characteristics of the multirotor UAV, and the second one placed below describes the consumed energy by the fixed-wing type.~The values presented in the figure were not normalized in the relation to the same value, but each one was scaled according to the maximum value obtained for a particular study case. Thus, lower consumption that can be observed for the multirotor does not mean this type of UAV uses less energy than the fixed-wing type. The main goal of the investigation was to highlight the impact of adding extra equipment to the UAV weight on the energy utilization itself in different seasons of the year. For a hovering multirotor UAV the amount of its total mass has a~crucial meaning to its energy demand.~By adding a PV panel and wind turbine (around $9$ kg) to the base station, its energy consumption increases almost twice. For different seasons, some deviations in the use of energy resources by the multirotor UAV can be observed as well. The reason for that is fluctuating air density value during the year. For the UAV of the fixed-wing type, aforementioned weight and weather dependencies can be noticed as well, but their impact on PC is not so significant as for the multirotor BS. On the other hand, within this paper, only one situation was considered, in which the fixed-wing UAV is flying with a~constant velocity equal to $10$ m/s.~The real impact might be seen by studying changes in energy utilization of the drone lifting various weight loads from the ground, or studying changes in its aerodynamic characteristics.~Both cases will be considered in future work.

\begin{figure}[t]
\label{figGeneration}
\centering
\includegraphics[width=0.49\textwidth]{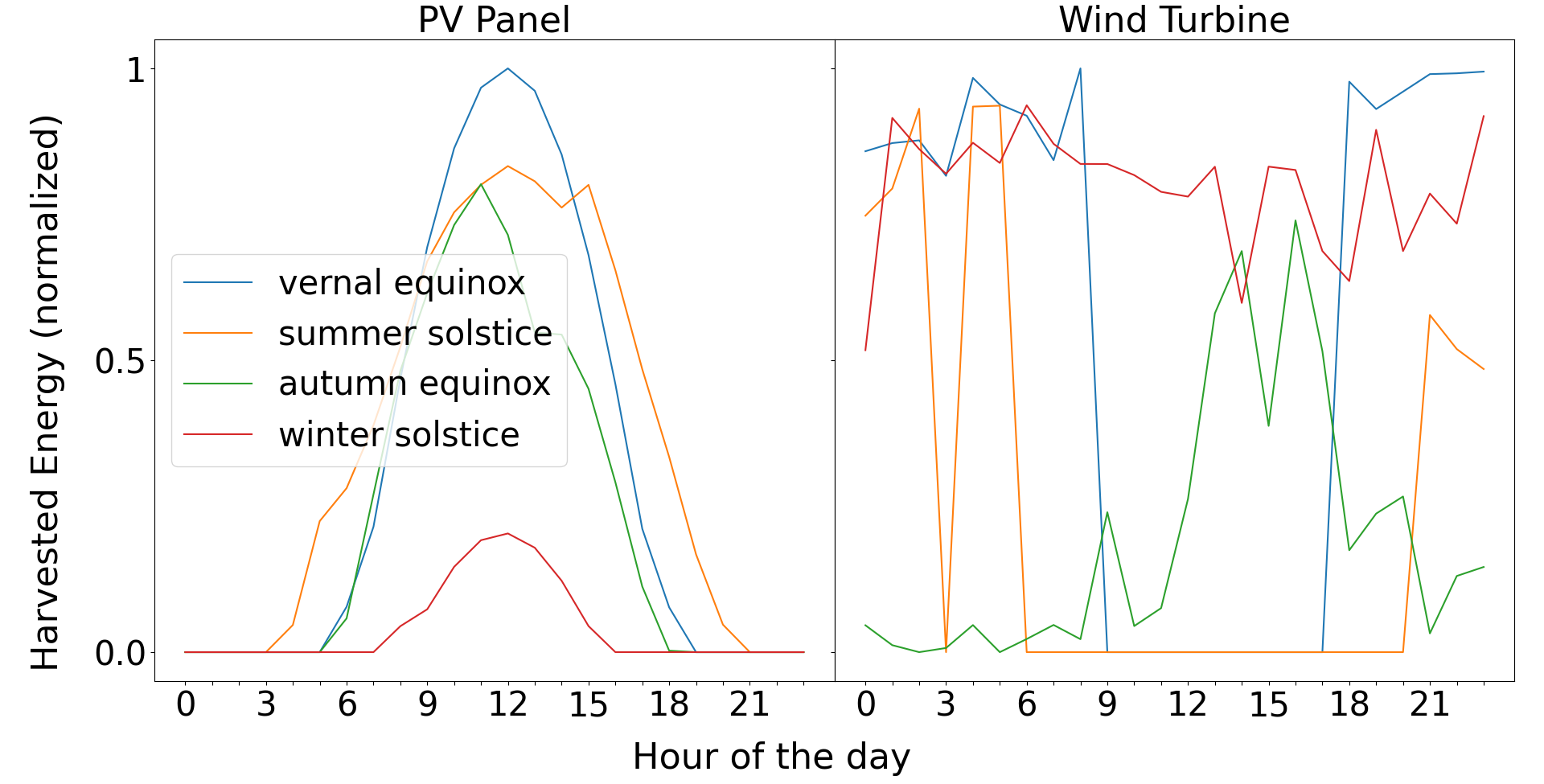}
\caption{Energy generation characteristics for PV panel and wind turbine.}
\end{figure}

Fig.~\ref{figGeneration} presents the energy harvesting processes of the PV panel and wind turbine used in the contribution to power the UAV BSs. The characteristics have been prepared based on the real equipment products, the specifications of which can be found in \cite{16} and \cite{17}.~The charts contained in Fig.~\ref{figGeneration} highlight the energy gain that can be achieved for different seasons of the year.~Based on that it is possible to assess when equipping the mobile base station with a particular RES generator is beneficial.~For instance, in the summer season, the PV panel is the most effective and WT the least. In the wintertime of the year, the situation is the opposite.~Interestingly, the peak of energy harvesting for both generators can be observed for the vernal equinox although it is not the most effective season for producing power by PV panels or wind turbines.~However, the reason for that is the limitations of RES generators, i.e., each type has its temperature or wind speed ranges, in which they can produce renewable energy resources.~In addition, the output power of the PV panel is greatly influenced by the current temperature of its solar cells (dependent, e.g., on the current ambient temperature) and the opacity of the clouds in the sky. Thus, an~increase in air temperature and the amount of clouds can cause a decrease in energy harvesting efficiency.

\section{Conclusions}
\label{secConclusions}
In this paper, the characteristics of energy utilization and generation by the two types of UAV mobile base stations equipped with RF transceivers, IRS devices, and RES generators (PV panels and wind turbines) have been investigated. Obtained results highlight the impact of the UAV's equipment on the energy balance for different seasons of the year. The paper's outcomes seem to be valuable for the process of network planning for future wireless systems considering the use of UAVs, IRSs, and RESs.


\end{document}